\documentclass[12pt]{iopart}

\usepackage{iopams}
\usepackage{setstack}

\usepackage{cancel}
\usepackage{float}
\usepackage{ifsym}
\usepackage{color}
\usepackage{latexsym}
\usepackage{latexsym}
\usepackage{dcolumn}
\usepackage{epsfig}

\usepackage{graphicx}

\newcommand{\bnull}{{\mathbf{0}}}

\newcommand{\be}{{\mathbf{e}}}

\newcommand{\bN}{{\mathbf{N}}}

\newcommand{\bx}{{\mathbf{x}}}
\newcommand{\by}{{\mathbf{y}}}

\newcommand{\al}{\alpha}

\newcommand{\la}{\lambda}

\newcommand{\sig}{\sigma}

\newcommand{\om}{\omega}
\newcommand{\Om}{\Omega}

\newcommand{\calH}{{\cal H}}

\newcommand{\calO}{{\cal O}}

\newcommand{\calZ}{{\cal Z}}

\newcommand{\wh}{\widehat}

\newcommand{\entwederoder}[4]{\left\{\begin{array}{lll}#1,&&#2\\#3,&&#4\end{array}\right.}

\begin{document}

\title[Microcanonical Determination of Interface Tension]{Microcanonical Determination of the Interface Tension of Flat and Curved Interfaces from Monte Carlo Simulations}
\author{A.~Tr\"oster and K. Binder}
\ead{troestea@uni-mainz.de}
\address{Johannes Gutenberg Universit\"at Mainz,
Staudingerweg 7,
D-55099 Mainz, Germany}

\date{\today}

\begin{abstract}
The investigation of phase coexistence in systems with multi-component
order parameters in finite systems is discussed, and as a generic example, Monte Carlo simulations of the 
two-dimensional q-state Potts model (q=30) on $L\times L$ square lattices ($40 \le L \le 100$) are presented.
It is shown that the microcanonical ensemble is well-suited both to find the precise location of the first order 
phase transition and to obtain an accurate estimate for the interfacial free energy between coexisting ordered 
and disordered phases. For this purpose, a microcanonical version of the heatbath algorithm is implemented.
The finite size behaviour of the loop in the curve describing the inverse temperature versus energy density is 
discussed, emphasizing that the extrema do not have the meaning of van der Waals-like "spinodal points" separating 
metastable from unstable states, but rather describe the onset of heterophase states: droplet/bubble evaporation/condensation 
transitions. Thus all parts of these loops, including the parts that correspond to a negative specific heat, describe  
phase coexistence in full thermal equilibrium. However, the estimates for the curvature-dependent interface tension of 
the droplets and bubbles suffer from unexpected and unexplained large finite size effects which need further study.
\end{abstract}


\maketitle

\section{Introduction}
In the theory of first order phase transitions,  a quantity of major interest is the interface tension
between phases in metastable coexistence. In particular, in the study of nucleation phenomena one faces the problem to 
determine the curvature dependence of interface free energy/entropy between a nucleating droplet or bubble 
of the emerging stable and its surrounding metastable phase
\cite{Oxtoby_JPCM4_1992,Katz_PAC64_1992,Debenedetti_MSL_1996,Kashchiev2000,Anisimov_RCR72_2003,BinderStauffer_AP25_1976,Binder_RepProgP50_1987}.
While an exact understanding of metastability in
infinite systems is currently still lacking from the point of view of rigorous statistical mechanics
\cite{PenroseLebowitz_JSP3_1971,Penrose_Lebowitz_1987},
one can nevertheless try to obtain valuable information from the study of phase separation in \emph{finite} 
systems \cite{BinderKalos_JStatPhys22_1980,FurukawaBinder_PRA26_1982,Binder_PhysicaA_2003}.
As analytical calculations are quite difficult for any nontrivial Hamiltonian, one resorts to simulations. 
For systems like Ising type spin models (lattice gases) \cite{WinterVirnauBinder_PRL103_2009}, 
binary mixtures \cite{Block_JCP133_2010} and simple fluids \cite{Schrader_PRE79_2009,Block_JCP133_2010}, that have been studied by 
simulation, a convenient scalar order parameter like the total magnetization
or particle number, which is extensive and even additive under partitions of the total system volume into subvolumes,
is available, and whose density serves to distinguish between the 
different phases. Phase coexistence in an equilibrated finite system is characterized by the identity of
the corresponding conjugate intensive quantity, whose physical meaning is that of an applied magnetic field or chemical potential.
Thus one can extract the order parameter bulk densities of the coexisting phases from analyzing 
free energies obtained from Monte Carlo \cite{BinderKalos_JStatPhys22_1980,FurukawaBinder_PRA26_1982,Schrader_PRE79_2009,Block_JCP133_2010} or 
molecular dynamics \cite{thompson:530,WoldeFrenkel_JCP109_1998,Kharlamov201110} simulations.
In particular, armed with an additive order parameter, one can determine the equimolar volumes of the coexisting phases
by evaluating the condition of vanishing adsorption of this order parameter for the corresponding choice of dividing surface.
From such data it is straightforward to compute the interface tension \cite{Schrader_PRE79_2009,Block_JCP133_2010}.
 
However, problems arise in cases for which a scalar additive order parameter is not available.
On the one hand, the different phases may be characterized by different values of a multi-component rather than a simple scalar quantity
(e.g.~the formal vector of particle numbers of different species in multi-component fluids).
On the other hand, there are cases where an explicit microscopic expression for an order parameter is frankly not known
like e.g.~in the study of protein folding.

An extensive parameter that is always on hand and whose density -- except for somewhat degenerate cases like hard spheres -- may be
utilized to distinguish different coexisting metastable and stable phases is the energy $E$. But unlike in the case of magnetization or
particle number, the notion of an ``equi-energetic'' surface seems to be completely counter-intuitive, as 
attributing zero energy to the interface is in conflict with the fundamental physical principle that interfaces are regions
with energy densities that are usually considerably larger than bulk ones. Thus, it is not clear how a reasonable measure of the 
corresponding subvolumes of the phases
can be obtained from a microcanonical analysis based solely on monitoring the energy density.  The purpose of the present
paper is to show how this can be done, and that energy may still be a ``good'' parameter to determine interface tensions. 

\section{Review of the Grand Canonical Route to the Interface Tension}

We start by reviewing the traditional Gibbs dividing surface approach to the description of phase coexistence 
of two phases labelled $\al$ and $\beta$ in a fluid
\cite{OnoKondo_HBdPh10_1960,RowlinsonWidom_1982,Debenedetti_MSL_1996,Kalikmanov_StatPhysFluids_2001,Vehkamaki2006}. 
Suppose first that the fluid has only a single chemical component. In spite of the fact that the problem of how to
rigorously define coexistence of a droplet of
phase $\al$ in (unstable) equilibrium with the surrounding $\beta$ phase has not been solved up to date 
for the case of an infinite system, it makes sense to consider
such a situation in a large but \emph{finite} system, in which such an equilibrium may actually be stabilized by
imposing appropriate thermodynamic boundary conditions.  
Phase separation into two connected regions is then usually detected by observing regions of different density
by monitoring the average inhomogeneous density profile $\rho(\bx)$.
Following Gibbs one may then choose an arbitrary dividing surface, defined as a levelling surface of zero thickness normal to 
the density gradient field, thus splitting the total volume $V$ into subvolumes
\begin{eqnarray}
V=V_\al+V_\beta.
\label{eqn:nnqklxmqklxmqklxmqklx}
\end{eqnarray}
Once this separation is agreed upon, any other extensive observable $M$ can be split into homogeneous and so-called excess contributions
as follows.
If $M$ assumes homogeneous equilibrium densities $m_\al,m_\beta$ in the phases $\al,\beta$, we set 
\begin{eqnarray}
M=M_\al+M_\beta+M^x,
\end{eqnarray}
where  
\begin{eqnarray}
M_\al\equiv V_\al m_\al,  \qquad M_\beta\equiv V_\beta m_\beta.  
\end{eqnarray}
In a similar way, we construct the excess energy $E^x$, entropy $S^x$, Helmholtz free energy $F^x$,
grand potential $\Om^x$, particle number $N^x$, and so on from their homogeneous densities and the division (\ref{eqn:nnqklxmqklxmqklxmqklx}). 
In a box of volume $V=L^d$, planar interfaces will usually form parallel to one pair of limiting walls.
The excess quantities defined above will then generally depend on the chosen position of the dividing surface with area $A=L^{d-1}$.
A notable exception is $\Om^x$, as can be understood from the fact that the corresponding homogeneous densities
are the negative pressures of both phases, which in case of a planar interface must agree by simple stability arguments. 
Thus, for a planar phase separation geometry, the interface tension
\begin{eqnarray}
\sig=\Om^x/A
\label{eqn:ndjswnqlwqlndwqlklndqndql}
\end{eqnarray}
is well defined, regardless of any parallel shift of the dividing surface. In contrast, the adsorption
\begin{eqnarray}
\Gamma=N^x/A 
\end{eqnarray}
changes with such a parallel shift of $A$. In turn, the condition of vanishing adsorption $\Gamma\equiv 0$
then uniquely fixes the position of $A$ and leads to an intuitively appealing definition of the ``actual'' position of the interface,
known as the \emph{equimolar surface}.

While it is straightforward to generalize the definition of the position of such an equimolar dividing surface 
from the planar to that of a spherical or otherwise curved case,
the definition of the corresponding interface tension now requires considerably more care. 
For a spherical interface in $d=3$ let us agree to use $\alpha$ to label the phase inside the 
spherical volume and $\beta$ to label the surrounding one outside the sphere.
To stabilize a curved interface, the pressures on both sides of the interface must necessarily be different.
In classical macroscopic physics this fact is encoded in the Laplace-Young (LY) equation
\begin{eqnarray}
\Delta p=p_\al-p_\beta=\frac{2\sig}{R},
\label{eqn:nxjdnodqkdxqqlnql}
\end{eqnarray}
which was derived from a mechanical analysis of the surface tension at the beginning of the 19th century.
When promoting the statistical mechanics definition (\ref{eqn:ndjswnqlwqlndwqlklndqndql}) from the planar case to that 
of curved interfaces, since $\Delta p\ne 0$ for $p_i=\Om_i/V=:-\om_i,i=\al,\beta$, the grand potentials densities $\om_\al,\om_\beta$ 
of both phases will also disagree and so 
\begin{eqnarray}
\Om^x=\Om^x(R)=V\om-V_\al(R)\om_\al-V_\beta(R)\om_\beta
\label{eqn:cnqjdnqjdnwqjnqjlnqjlxdnqj}
\end{eqnarray}
is bound to pick up a dependence on the radius. 
Thus, for a curved interface the interface tension $\sig=\sig(R)$ appearing in (\ref{eqn:nxjdnodqkdxqqlnql}) will itself be $R$-dependent.
At this point, however, it is crucial to realize that within the Gibbs dividing surface approach the choice of the radius $R$, being a purely theoretical construct,
is in principle arbitrary, such that the classical 
relation (\ref{eqn:nxjdnodqkdxqqlnql}) can only hold for a distinguished value of $R$.
Nevertheless, physically observable quantities should not depend on the position of the artificially introduced dividing surface.
Indeed, differentiating the definition of the excess grand potential in d dimensions with respect to $R$ while keeping all other variables fixed,
which is called a \emph{notional derivative} and indicated by a bracket notation $[d/dR]$, one arrives at the \emph{generalized LY equation}
\cite{OnoKondo_HBdPh10_1960,RowlinsonWidom_1982}
\begin{eqnarray}
\Delta p=(d-1)\frac{\sig(R)}{R}+\left[\frac{d\sig(R)}{dR}\right].
\label{eqn:nxjdnodqkdxqqlnqlgeneralized}
\end{eqnarray}
The classical LY equation (\ref{eqn:nxjdnodqkdxqqlnql}) may only be recovered from this equation for
the special choice of radius $R=R_s$, for which
\begin{eqnarray}
\left[\frac{d\sig(R)}{dR}\right]_{R=R_s}\equiv 0.
\label{eqn:hbebebeekkeekxc}  
\end{eqnarray}
To account for this fact, the corresponding dividing surface is known as the \emph{surface of tension}.
Thus, $R_s$ is a stationary point of $\sig(R)$  under a notional variation of $R$.
In fact, it is not hard to see \cite{OnoKondo_HBdPh10_1960,RowlinsonWidom_1982} that $R_s$ is indeed a minimum of $\sig(R)$, which can explicitly be
deduced from the general form of $\sig(R)$ in $2$ or $3$ dimensions
\begin{eqnarray}
\frac{\sig(R)}{\sig(R_s)}=1+\entwederoder{\frac{1}{2}\left(\frac{R-R_s}{R}\right)^2\frac{R}{R_s}}{d=2}
{\frac{1}{3}\left(\frac{R-R_s}{R}\right)^2\frac{(R_s+2R)}{R_s}}{d=3},
\label{eqn:dddqdwqdhwqdwqbdwqbdwqbd}
\end{eqnarray}
according to which $\sig(R)$ is universally determined from knowledge
of $R_s$ and $\sig_s\equiv\sig(R_s)$.  The difference
$\delta(R_s):=R_e-R_s$ between radius of the equimolar surface $R_e$
and $R_s$, which has become famous under the name \emph{Tolman
  length}, is known to be of molecular sizes.  Of course, in principle
one could also compute physical observables from e.g.~the equimolar
interface tension $\sig_e\equiv\sig(R_e)$ or any other choice of
radius $R$, since all physical information is encoded in any such
choice. However, the choice $R_s$ is particularly convenient, as the
condition (\ref{eqn:hbebebeekkeekxc}) allows to condense many formulas
to a considerable more compact and manageable form. 
To some extent this is also true for the choice
$R=R_e$, but the definition $\Gamma_e=\Gamma(R_e)\equiv 0$ explicitly
makes use of the particle number $N$ as an extensive parameter.  For a
multi-component fluid of $\tau>1$ different chemical components, which
we may label by an index $t=1,\dots,\tau$, this complicates manners in
a considerable way. In fact, the particle number $N$, adsorption
$\Gamma$, associated chemical potential $\mu$ are then both promoted
from scalar quantities to formal vectors $\bN$, $\bGamma$,
$\bmu$, where $\bN=(N^1,\dots,N^\tau)$ and so on.  One now must
deal with $\tau$-component averaged density profiles
$\brho(\bx)=(\rho^1(\bx),\dots,\rho^\tau(\bx))$.  In general,
such systems have complex phase diagrams, and if we concentrate on a
particular transition, the levelling surfaces of different density
components $\rho^i(\bx)$ may yield different equimolar surfaces for
each chosen component, relative to which the remaining components form
inhomogeneous adsorption layers. In principle it may still be possible to
eliminate clumsy adsorption terms from formulas to some extent by an
ingenious choice of dividing surface, the so-called Koenig dividing
surface \cite{Boltachev2003228}. However, to fulfil its defining
condition, all $\tau$ components of the adsorption have to be balanced
simultaneously with the multiple components of the associated chemical
potential, which leads to profound numerical difficulties in the
evaluation of simulation results.


To illustrate these calamities, let us review the practical steps to calculate the interface tension from
Monte Carlo free energy simulation data for a one component fluid and compare them to the expected effort for
a multi-component one.
We choose a cubic simulation box of size $N=L^d$ with periodic boundary conditions. Carrying out
grand canonical Monte Carlo simulations at some fixed temperature $T_0$ chosen somewhat lower than the critical temperature $T_c$,
one first has to determine the coexistence chemical potential $\mu=\mu_0(T_0)$. Practically, this is done by implementing the
equal--weight rule \cite{BorgsKotecky_JStatPhys61_1990} numerically by performing 
a histogram re-weighting \cite{FerrenbergSwendsen_PRL23_1988} to the simulated grand canonical probability distribution $P_{T_0V\mu}(N)$
of particle numbers and extrapolating the result to the thermodynamic limit $L\to\infty$. For a multiple component fluid,
these steps are already quite involved, as the approximately Gaussian-shaped peaks of $P_{T_0V\bmu}(\bN)$ are defined on a multidimensional space of
variables and in presence of $q$ possible low temperature phases one has to deal with $q+1$ of them in order to pin down
the $\tau$ individual components of the vectorial chemical potential $\bmu$ to their coexistence values $\bmu_0$.

Nevertheless, suppose that this task has been carried out successfully. In the single component case, one
next needs to resolve the detailed structure of $P_{T_0V\mu_0}(N)$ by e.g.~Wang-Landau sampling, eliminating residual inaccuracies
by a weighted Monte Carlo production run. In this way, one obtains a dimensionless excess free energy density
\begin{eqnarray}
-\beta \hat f^{(L)}(T_0,\rho)\equiv (1/V)\ln P_{T_0V\mu_0}(N),  
\label{eqn:cnjcnoqcmqkcnqjlcnqjlcnqjlcnqjcnql}
\end{eqnarray}
which depends on the scalar density $\rho=N/V$ (we have dropped a purely $T_0$-dependent normalization part which is irrelevant for what follows;
the usual thermodynamic notation $\beta_0=1/k_BT_0$ should not lead to any confusion with the label $\beta$ for one of the two phases).
At a first order phase transition, the finite size potential $\beta_0 \hat f^{(L)}(T_0,\rho)$ has a double-welled shape
with a flat central plateau, and an analysis of its fine details reveals several distinct ranges of
the total density $\rho$ between its two minima, for which one may observe phase separation in planar, cylindrical (in $d=3$) and spherical
shapes induced by the imposed periodic boundary conditions. These regions are also detectable in the derived
finite size \emph{canonical} ``excess'' chemical potential 
\begin{eqnarray}
\hat\mu^{(L)}(T_0,\rho)=\left(\frac{\partial\hat f^{(L)}(T_0,\rho)}{\partial \rho}\right)_{T_0}.
\label{eqn:xnxcjcnqxnqxnnbnbkqbdnqjkxdnqjkxdnqk} 
\end{eqnarray}
From the distorted ``Van der Waals loop'' shape of $\hat\mu^{(L)}(T_0,\rho)$, one can infer that
for small enough $\hat\rho$ the equation $\mu^{(L)}(T_0,\rho)\equiv \mu$ generally allows for at least three roots 
$\rho^{(L)}_\al(T_0,\mu)<\rho^{(L)}(T_0,\mu)<\rho^{(L)}_\beta(T_0,\mu)$
corresponding to two bulk densities $\rho^{(L)}_\al,\rho^{(L)}_\beta$ of coexisting phases $\al,\beta$ at total density $\rho$.
This allows to introduce the total dimensionless grand potential density 
\begin{eqnarray}
\om^{(L)}(T_0,\mu)=\wh f^{(L)}(T_0,\rho(T_0,\mu))-\mu\rho^{(L)}(T_0,\mu)
\label{eqn:cnjcnnnlqnllxnkkwkwnwnlns}
\end{eqnarray}
and the grand potential densities
\begin{eqnarray}
\om^{(L)}_\al(T_0,\mu)&=&\wh f^{(L)}(T_0,\rho_\al(T_0,\mu))-\mu\rho^{(L)}_\al(T_0,\mu),  \\
\om^{(L)}_\beta(T_0,\mu)&=&\wh f^{(L)}(T_0,\rho_\beta(T_0,\mu))-\mu\rho^{(L)}_\beta(T_0,\mu),  
\label{eqn:xdjxmklxmkxmkxmxm}
\end{eqnarray}
from which it is straightforward to compute $\Om^x$ for a given volume partition $V=V_\al(R)+V_{\beta}(R)$,
and thus, using (\ref{eqn:ndjswnqlwqlndwqlklndqndql}), the notionally $R$-dependent interface tension $\sig^{(L)}(R)$.
In detail, $R_s$ and $\sig^{(L)}_s$ are found by numerical minimization of $\sig(R)$ with respect to $R$,
while $\sig^{(L)}_e$ can be calculated from the spherical equimolar volume determined by the lever rule
\begin{eqnarray}
\frac{V_\al(R_e)}{V}=\frac{\rho-\rho_\al}{\rho_\beta-\rho_\al},\quad 
\frac{V_\beta(R_e)}{V}=\frac{\rho_\beta-\rho}{\rho_\beta-\rho_\al},
\label{eqn:xnxcnxcnxnxcnsjlxnsqjlxnqnxsqx}
\end{eqnarray}
for $V_\al(R)=4\pi R^3/3$ in $d=3$ and $V_\al(R)=\pi R^2$ in $d=2$, respectively.
In passing we note that since $\Om^x(R_e)=F^x(R_e)$,
$\sig_e^{(L)}$ is exclusively determined by $\wh f^{(L)}(T_0,\rho(T_0,\mu))$ (cf.\ \cite{Schrader_PRE79_2009,Block_JCP133_2010}).

All this is very fine -- however, in a multi-component setting the numerical effort to 
carry out these steps successfully is again forbidding. What is needed is an intrinsically scalar approach.

\section{Microcanonical Approach}

As mentioned above,  a basic extensive scalar quantity that is always available is the total energy $E$ of the system.
It is by now well known that while phase transitions are 
only well defined in an infinite system in the strict sense, they can nevertheless 
be studied conveniently by analyzing the so-called ``convex intruder'' in the microcanonical entropy $S(E)$ of a finite version of the system
\cite{Thirring_ZFP235_1970,Gross_MCT_2001}.
\begin{figure}[tb]
\centering
\includegraphics[scale=0.35]{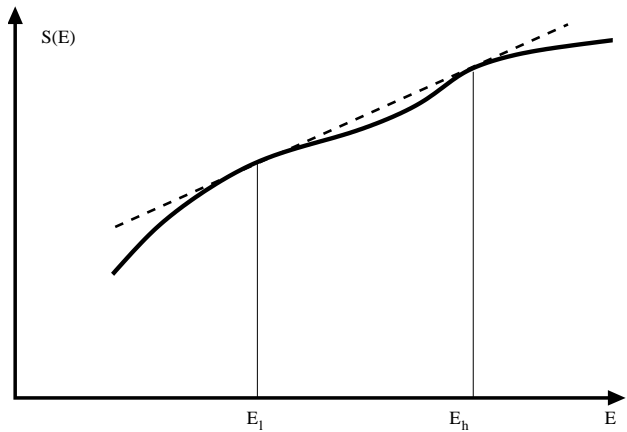}
\caption{Schematic illustration of a composite entropy with a convex intruder.}
\label{fig:iiibbfehbhebhfwevgwevwvw}
\end{figure} 
By its very definition, such a convex intruder in $S(E)$ gives rise to a corresponding anomaly of the microcanonical inverse temperature $\beta(E)$
and a possibly negative branch of the accompanying microcanonical specific heat with quasi-singularities at the intruder boundaries
(for a schematic picture, see Figure \ref{fig:iiibbfehbhebhfwevgwevwvw} below).
In a finite (or at least in some sense ``small'' \cite{Thirring_ZFP235_1970})
system such an observation does not contradict thermodynamic stability requirements \cite{Callen_TIT_1985}.
One now observes that up to a sign, the overall appearance of $\beta(E)$ is quite similar to that of the excess chemical potential
$\hat\mu^{(L)}(\beta,\rho)$ discussed above. Indeed, parallels between the former constructions and the ones carried below are not accidental. 

Recording the energy probability distribution $P_{T_0V\mu}(E)$ at fixed $T_0$ for different values of $\mu$, we can 
locate the coexistence chemical potential $\bmu_0=\bmu_0(T_0)$ by applying the equal weight rule 
\cite{ChallaLandauBinder_PRB34_1986,BorgsKotecky_JStatPhys61_1990,BorgsKotecky_PRL68_1992}
to the two separate approximately Gaussian-shaped peaks, as will be discussed below for the case of the Potts model.
In any case, this may be feasible even in case of a multi-component order parameter, since the domain of $P_{T_0V\bmu}(E)$ is the scalar quantity $E$.
For the rest of the simulation, the parameters $T_0$ and $\bmu_0$ remain fixed.
It is then convenient to formally replace the energy $E$ by a grand-canonical version
\begin{eqnarray}
E-\bmu_0\bN  \ \to\ E
\end{eqnarray}
in the same way as one may formally replace the original Hamiltonian by a grand-canonical one in computing the grand-canonical partition function. 
For our purposes, the ``background'' homogeneous chemical potential $\bmu_0$ can thus be eliminated from the list of thermodynamic variables.
With $\bmu_0$ fixed, we now record the number of states 
\begin{eqnarray}
g(E,V)\equiv e^{S(E,V)}
\end{eqnarray} 
with grand canonical energy $E$ in a flat histogram simulation followed by a weighted Monte Carlo production run. 
Once $g(E,V)$ is known, we arrive at the grand canonical partition function in the form
\begin{eqnarray}
\calZ(\beta,V)=\sum_{E}e^{-\beta E+S(E,V)}  
=\sum_{e}e^{-V(\beta e+s^{(L)}(e))},  
\end{eqnarray}
where we introduced the energy and entropy densities $e=E/V$ and $s^{(L)}(e)=S(E,V)/V$,
valid for arbitrary $\beta$.
Suppose that at a particular inverse temperatures $\beta$ the sum is dominated by its largest summand
in the limit of large system size. The corresponding energy density $e(\beta)$ is determined by the equation
\begin{eqnarray}
\beta\equiv\left(\frac{\partial s^{(L)}(e)}{\partial e}\right)_V\Bigg|_{e=e^{(L)}(\beta)}  =\beta^{(L)}(e)|_{e=e^{(L)}(\beta)},
\label{eqn:msqkmqxmqkxmqxmqxmq}
\end{eqnarray}
which expresses the equality of inverse canonical temperature $\beta$ and microcanonical temperature $\beta^{(L)}(e)$
for this special value of $e$.
One may then approximate $\beta\Om(\beta,V,\bmu_0)=-\log\calZ(\beta,V,\bmu_0)$ by
\begin{eqnarray}
\beta\om^{(L)}(\beta,\bmu_0)\approx \beta e^{(L)}(\beta)-s^{(L)}(e^{(L)}(\beta)), 
\end{eqnarray}
i.e.~as a Legendre transform. 
This equation is quite similar to (\ref{eqn:cnjcnnnlqnllxnkkwkwnwnlns}), $-s^{(L)}(e)$ playing the role of
$\hat f^{(L)}$, $e$ that of $\rho$ and $-\beta$ that of $\mu$. 
Similar to the strategy employed above for the chemical potential, for a certain range of inverse temperatures $\beta$
the equation $\beta\equiv \beta^{(L)}(e)$ has (at least) three different solutions $e_\al<e<e_\beta$ 
corresponding to the total energy density $e$ and those of the two coexisting phases $\al,\beta$. 
If we now let $\beta$ approach this interval around $\beta_0$, 
the presence of the convex intruder in $s^{(L)}(e)$ makes the above discrete saddle point approximation
break down since not one but at least three ``saddle points'' are found, which in the non-degenerate case of three roots correspond
to the dimensionless grand potential densities
\begin{eqnarray}
\beta\om^{(L)}(\beta)&=&\beta e-s^{(L)}(e), \label{eqn:a_ndebnefwewefwebf}\\ 
\beta\om^{(L)}_\alpha(\beta)&=&\beta e_\al-s^{(L)}(e_\al),\label{eqn:b_ndebnefwewefwebf}\\
\beta\om^{(L)}_\beta(\beta)&=&\beta e_\beta-s^{(L)}(e_\beta).\label{eqn:c_ndebnefwewefwebf}
\end{eqnarray}
At this stage, we have reached our goal announced above, as these quantities, once they are determined, allow
to compute the excess grand potential and thus the interface tension $\sig^{(L)}(R)$ from formula (\ref{eqn:cnqjdnqjdnwqjnqjlnqjlxdnqj}).
By minimization of $\sig^{(L)}(R)$ w.r.t~$R$ we obtain the radius of the surface of tension $R^{(L)}_s$ and the corresponding surface tension
$\sig^{(L)}_s$. A calculation of $R_e$ is, of course, not feasible in this approach.

\section{The 2d Potts Model}

We illustrate our strategy taking the example of the $q=30$ nearest neighbour Potts model in $d=2$.
The $q$-state Potts model \cite{Wu_RevModPhys.54.235}, whose Hamiltonian on a simple cubic 2d lattice of $N=L^2$ sites in zero external field is given by
\begin{eqnarray}
\calH[\{s(\bx)\}]=\sum_{\langle\bx\by\rangle}[1-\delta_{s(\bx),s(\by)}],  
\end{eqnarray}
where $s(\bx)\in\{1,\dots,q\}$ is ideally suited for this purpose for a number of reasons.
(i) First of all, for $q>4$ the model undergoes a temperature-driven first order phase transition. 
(ii) Regarding this transition, a wealth of rigorous results \cite{Wu_RevModPhys.54.235,Kihara_JPhSocJap_1954,Baxter_JPC_1973,Janke_PRB47_1993,BorgsJanke_JdPhI_1992}
is available in the literature which can serve to benchmark our simulation results. 
For the first order phase transition temperature of a bulk system, one has the exact analytic expression
\begin{eqnarray}
1/T_0=\beta_0 =\ln(1+\sqrt{q}).
\end{eqnarray} 
In \cite{BorgsJanke_PRL68_1992,Janke_PRB47_1993}
it was reported that, as was expected from general arguments,
the inverse temperature $1/T_0(L)=\beta_0(L)$ at which the ratio of the two weights of the thermal energy probability 
distribution  is just $q$, agrees with the exact bulk value $\beta_0\equiv \beta_0(L=\infty)$ up to exponentially small corrections. 
Thus, $\beta_0(L)$ serves as a convenient definition of a finite size transition temperature. 
Other rigorous results include the latent heat per volume \cite{Baxter_JPC_1973},
the limiting internal energy densities at $\mathbf{T_0}$ 
and the difference of specific heats \cite{Kihara_JPhSocJap_1954}.
Furthermore, the reduced interface tension (i.e.~the interface free energy density at $\beta_0^{-1}$ multiplied by $\beta_0$) 
between the disordered and one of the ordered phases
along the square lattice $(10)$ direction was rigorously determined \cite{BorgsJanke_JdPhI_1992} 
to be
\begin{eqnarray}
2\sigma_{\mathrm{o/d}}=4\sum_{n=0}^{\infty}\ln\frac{1+w_n}{1-w_n},
\label{eqn:bnchebeidqednqjdqddbnqw}
\end{eqnarray}
where
\begin{eqnarray}
w_n:=\left(\sqrt{2}\cosh\frac{(n+1/2)\pi^2}{2v}\right)^{-1}  
\end{eqnarray}
with
\begin{eqnarray}
v:=\ln\left[\frac{1}{2}\left(\sqrt{\sqrt{q}+2}+\sqrt{\sqrt{q}-2}\right)\right]. 
\end{eqnarray}
Equally important for us is the fact that there is even a rigorous calculation of the full anisotropic interface tension available
\cite{Fujimoto_JPhysA_1997}. However, since these calculations are too 
involved to be reproduced here, we content ourselves with noting that at $\beta_0$ the resulting anisotropy for $q=30$ calculated from the formulae
in \cite{Fujimoto_JPhysA_1997} is vanishingly small. This happenstance is a very important prerequisite for any attempt to 
apply our evaluation strategy for the interface tension, which rests on a presupposed spherical symmetry of bubbles and droplets.

(iii) The $q$ state Potts model's order parameter is not scalar, but has dimension $q-1$.
Since this is a central issue in the present context, let us briefly review its nuts and bolts.
Guided by physical intuition, a scalar ``order parameter'' $m$ could be defined by the following reasoning
\cite{Wu_RevModPhys.54.235,ChallaLandauBinder_PRB34_1986}.
Let  $N^{(a)}$ denote the number of spins of a given microstate with value $s(\bx)=a$, where $1\le a\le q$.
Let $N_{\mathrm{max}}:=\max(N^{(1)},\dots,N^{(q)})$. Then
\begin{eqnarray}
M:=\frac{q\langle N_{\mathrm{max}}\rangle-N}{q-1}.
\label{eqn:jnjnxjnxlnlnlnsnlnlsnlsnls}
\end{eqnarray}
Obviously $m:=M/N$ is confined to values $0\le m\le 1$, and $m=0$ for complete disorder, while $m=1$ for any perfectly ordered domain.
In the Ising case $q=2$, $m$ indeed corresponds to the \emph{modulus} of the magnetization density of the system.
Thus, if we break up the system volume into subvolumes $V_i$ and add up their different $M_i$'s, 
generally  $M\ne \sum_iM_i$, i.e.~$M$ is \emph{not} additive between subsystems. 

On the other hand, in \cite{ZiaWallace_JPhysA_1975} Zia and Wallace construct a full $(q-1)$-component order parameter.
They introduce $q$ unit vectors in $\be^{(a)}\in\mathbb{R}^{q-1},\ a=1,\dots, q,\ q>1$, such that the following relations are satisfied:
\begin{eqnarray}
\be^{(a)}\be^{(b)}=\frac{q\delta^{ab}-1}{q-1}.
\label{eqn:whbheqbdqbdqhbqhbcqhbcqjhvgqvgqvgqvxdgq}
\end{eqnarray}
Any set of such vectors defines a generalized tetrahedron in $\mathbb{R}^{q-1}$,
the $q$ vectors pointing from the center to each corner. Trivially, the vectors $\be^{(a)}$ cannot be linearly independent.
Instead, they satisfy the geometrically evident sum rule
\begin{eqnarray}
\sum_{a}\be^{(a)}=\bnull.  
\label{eqn:xqhbhbhwjhwkbwkbwkbwks}
\end{eqnarray}
With the one-to-one correspondence $s(\bx)\ \Leftrightarrow \ \be^{(s(\bx))}=:\be(\bx)$ understood, 
one associates a \emph{$q-1$ component order parameter}
\begin{eqnarray}
\mathbf{M}:=\sum_{\bx\in\Gamma}\be(\bx)\ \in\mathbb{R}^{q-1}
\end{eqnarray}
with each given microstate, which we will call the \emph{magnetization}. 
In terms of the occupation numbers $N^{(a)}$ of the Potts spin states
\begin{eqnarray}
\mathbf{M}=\sum_{a=1}^q N^{(a)}\be^{(a)}.
\label{eqn:dnjxdnnjkwnwjksnwnswjnsw1j}
\end{eqnarray}
If one multiplies (\ref{eqn:dnjxdnnjkwnwjksnwnswjnsw1j}) with any of the unit vectors $\be^{(b)}$, it is easy to see that
\begin{eqnarray}
M^{(b)}:=\be^{(b)} \mathbf{M}=\frac{qN^{(b)}-N}{q-1}.
\label{eqn:eqbqhkbdwjkndwjnnjlns}
\end{eqnarray}
Clearly, $m^{(b)}:=M^{(b)}/N\in\left[-1/(q-1),1\right]$ agrees with $m$ as defined in (\ref{eqn:jnjnxjnxlnlnlnsnlnlsnlsnls}) 
provided $N^{(b)}=N_{\mathrm{max}}$. This clarifies the role of $m$ as well as the low temperature domain structure of the model.
Namely, suppose that $N^{(b)}=N_{\mathrm{max}}$. Then, we can rewrite (\ref{eqn:dnjxdnnjkwnwjksnwnswjnsw1j}) as
\begin{eqnarray}
\mathbf{M}=\sum_{a\ne b} \underbrace{(N^{(b)}-N^{(a)})}_{\ge0}(-\be^{(a)}).
\end{eqnarray}
The $q$ domains $D^{(b)}$ are thus geometrically represented by convex cones enclosed by the set of vectors
$\{(-\be^{(a)}): a=1,\dots,q,a\ne b\}$, and (\ref{eqn:eqbqhkbdwjkndwjnnjlns}) gives the projection
of $\mathbf{M}$ onto the average direction $\sum_{a\ne b}(-\be^{(a)})=\be^{(b)}$,
the symmetric group of permutations of $q$ numbers acts as the underlying symmetry.
  
At small values of $m$, several very small fluctuation ``clusters'' 
of the same or competing spin values may coexist, causing the system to jump randomly from one domain cone to another.
On the other hand, the parameter $m^{(b)}$ is always additive by construction, but does depend on the chosen direction
$b$ in $\mathbf{M}$-space.   
Only for values of $m^{(b)}$ larger than a certain threshold do we find  agreement
of the order parameters computed from (\ref{eqn:jnjnxjnxlnlnlnsnlnlsnlsnls}) and (\ref{eqn:eqbqhkbdwjkndwjnnjlns}), 
since then the direction along which the projection
from $\mathbf{M}$ to $M$ occurs is uniquely determined by the value  $b$ of the majority occupation number. 
A cluster decomposition performed during the course of a simulation can provide 
information in analyzing these fluctuations in order parameter topology.
   
Thus, the parameter (\ref{eqn:jnjnxjnxlnlnlnsnlnlsnlsnls}) is only additive for magnetizations $\mathbf{M}_1$ and $\mathbf{M}_2$ 
both belong to a single common domain $D^{(b)}$, and thus cannot be used meaningfully as a parameter 
in a Gibbs dividing surface construction.
On the other hand, sampling the free energy as a function of the full order parameter
$\mathbf{M}$ is out of the question. In other words, we are exactly in the situation outlined 
earlier, and thus embark on a microcanonical strategy instead. Well, not quite. Actually, the situation
for the Potts model is not as complicated as the one outlined in the introductory section.
This is largely due to the fact that there is no need to determine a coexistence ``chemical potential'',
which would correspond to an external vectorial magnetic field. As the different $q$-spins of the Potts model
are all coupled in the same way, this external field is fixed to be exactly zero.

\begin{figure}[tbh]
\centering
\includegraphics[scale=0.25]{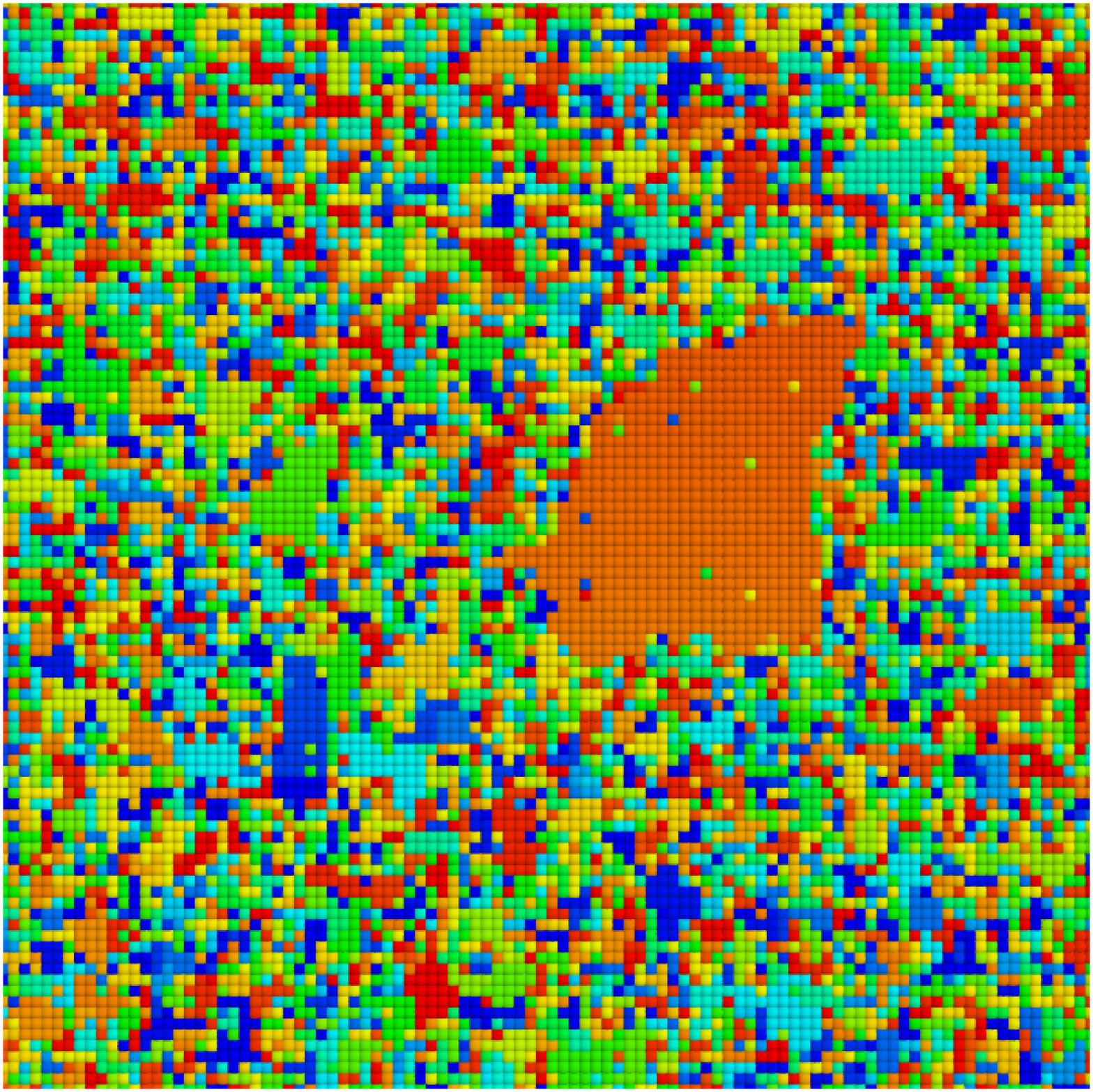}
\caption{$q=30$ Potts model at in 2 dimensions with periodic boundary conditions at $L=100$:
snapshot of typical ``droplet'' configuration at scalar order parameter value $m\approx 0.06$.}
\label{fig:delddmedmekmdekmekdmekm}
\end{figure}

\begin{figure}[tb]
\centering
\includegraphics[scale=0.25]{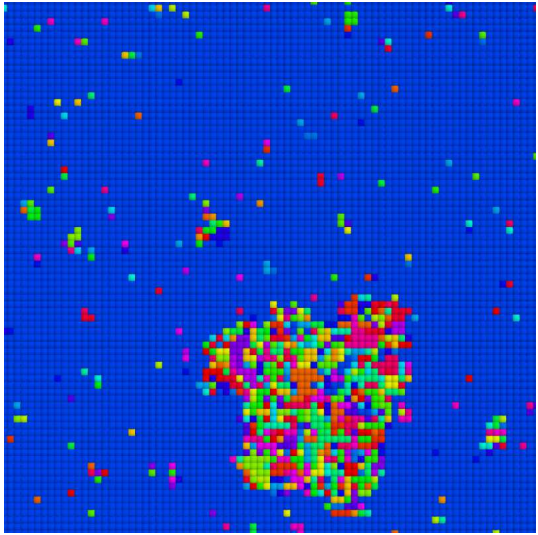}
\caption{$q=30$ Potts model in 2 dimensions with periodic boundary conditions at $L=80$:
snapshot of typical ``bubble'' configuration at scalar order parameter value $m\approx 0.87$.}
\label{fig:delddmedmekmdekmekdmekmsnnsnlsn}
\end{figure}

\section{The Microcanonical Heat Bath Algorithm}

Any successful design for a Monte Carlo algorithm devoted to the study of 
phase separation at phase transitions of pronounced first order character  
must address the phenomenon of exponentially diverging autocorrelation times
\begin{eqnarray}
\tau(L,\beta)\sim \exp(2\beta\sig_{\infty} L^{d-1})  
\label{eqn:xkmxdmqkxdmqomoqoq}
\end{eqnarray}
accompanying first order phase transitions which known as \emph{hyper-critical
slowing down} \cite{Weigel_Phys_Proc_2010} (in (\ref{eqn:xkmxdmqkxdmqomoqoq}) a $d$-dimensional cubic box of 
volume $L^d$ with periodic boundary conditions is assumed).
A Wang-Landau type of algorithm is capable of overcoming large 
entropy or free energy barriers separating different stable or metastable phases \cite{NeuhausHager_JStatPhys_2003}
and is in principle straightforward to implement.  
In its original microcanonical version, the algorithm directly yields the density of states (or, for a discrete system, rather number of states)  
\begin{eqnarray}
g(E)=\sum_{\nu}\delta(E_{\nu}-E)
\end{eqnarray}
of a system having microstates $\nu$ with energies $E_{\nu}$ (we are somewhat casual about the use of the Dirac delta function for simplicity, 
and we have put $k_B=1$ for the same reason) with high precision,
which is related to the microcanonical entropy by $S(E)=\ln g(E)$.
To control possible residual errors,
one may thus determine an approximate microcanonical density of states by Wang-Landau simulations and perform 
subsequent biased Monte Carlo production runs with statistical weights based on this approximate density of states. In fact,  
knowledge of $S(E)$ conveniently allows to determine a multitude of other $T$-dependent observables at various temperatures simultaneously 
with high precision, as is explained in more detail below.

It remains to construct a suitable move set for a microcanonical Wang-Landau simulation scheme.
Single $q$-spin updates are simple to implement and may be a reasonable choice for Ising systems far from criticality, 
but are quite inefficient in exploring
the regions of phase space of the large $q$ Potts model which are of interest for studying phase separation, 
namely those configurations, in which 
a single ordered domain of Potts spins $s(\bx)$ of, say, value $s(\bx)=q$ coexists 
with a disordered background. Indeed, suppose that during the course of the simulation, 
our random walk arrives at a particular configuration,   
in which almost all Potts spins agree with this condition, while just a few, say, $s(\bx_i),i=1,\dots,k$ are yet disordered.
In such a microstate, chances are only $N^{-1}\times (q-2)^{-1}$ that of the ``missing'' sites $s(\bx_i)$ is indeed drawn 
and its spin $s(\bx_i)$  be assigned the ``right'' value $q$ in creating the next trial configuration,
thus matching the surrounding domain. 

Within the canonical ensemble, it is well known that the \emph{heat bath algorithm}
\cite{LandauBinder_MC_2000,NewmanBarkema_MCMSP_1999} is superior to the standard Metropolis scheme for high $q$ Potts models.
In detail, let $E^{(\mu)}$ denote the total energy of the system in the microstate $\mu$.
Choose a random site $\bx$ and let $q_\mu:=s(\bx)$ denote the value of the Potts spin variable 
at this particular site. Furthermore, let
$\by_i,i=1,\dots,z$ denote the nearest neighbour sites of $\bx$.
Defining the \emph{local energy at site $\bx$} by
\begin{eqnarray}
E_q\equiv E_{\mathrm{(local)}}(q,\bx):= -\sum_{i=1}^z \delta_{q,s(\by_i)}, 
\end{eqnarray}
one can split
\begin{eqnarray}
E^{(\mu)}=E_{(\mathrm{local})}(q_\mu,\bx)+E^{(\mu)}_{\mathrm{rest}}
\end{eqnarray}
and define a set of heat bath probabilities
\begin{eqnarray}
p_q:=\frac{e^{-\beta E_q}}{\sum_{n=0}^{q-1}e^{-\beta E_n}},  
\end{eqnarray}
which are manifestly \emph{independent} of the value of the initial central Potts spin $s(\bx)$.
The canonical heat bath algorithm amounts to choosing a new value $q_\nu\in\{0,\dots,q-1\}$ for
this spin with probability $p_{q_\nu}$ in every step.
It is easy to see that the resulting algorithm satisfies detailed balance as well as ergodicity.
In terms of generation and acceptance probabilities, we have $g(\mu\to\nu)=\frac{p_{q_\nu}}{N}$ and $a(\mu\to\nu)\equiv 1$,
i.e.~the stochastic character only enters in the generation of configurations, which are, once generated, always accepted. 

It is straightforward to translate these ideas from the canonical to the microcanonical
setting. To illustrate the correspondence, let us denote the canonical Boltzmann weights by
\begin{eqnarray}
\pi_\nu:=  \frac{e^{-\beta E^{(\nu)}}}{Z(\beta)}.
\end{eqnarray}
Then the canonical heat bath probabilities $p_q$ can quite trivially be rewritten as
\begin{eqnarray}
p_{q_\nu}:=\frac{e^{-\beta (E_q+E_{\mathrm{rest}})}}{\sum_{n=0}^{q-1}e^{-\beta (E_q+E_{\mathrm{rest}})}}
=\frac{\pi_{q_\nu}}{\sum_{n=0}^{q-1}\pi_{n}}.  
\end{eqnarray}
Now, to translate the algorithm to the microcanonical ensemble, we simply
replace $\pi_\nu\to 1/g(E_\nu)$. 
In terms of the microcanonical entropy $S(E):=\ln g(E)$, we can rewrite the above probabilities as 
\begin{eqnarray}
p_{q_\nu}=\frac{1/g(E_\nu)}{\sum_{n=0}^{q-1}1/g(E_n)}  =\frac{e^{-S(E_\nu)}}{\sum_{n=0}^{q-1}e^{-S(E_n)}}. 
\end{eqnarray}
Once we have determined the microcanonical entropy $S(E)$, it is in principle straightforward to 
obtain the free energy as a function of $T$ and some other (preferable scalar) observable $o=\calO[\{s(\bx)\}]$,
where $\calO[\{s(\bx)\}]$ denotes e.g.~the magnetization $m$, the projection of the order parameter $m^{(a)}$ along
an arbitrary fixed internal direction $a$, the size of the largest geometric or Swendsen-Wang cluster, all 
simultaneously computed for different temperatures from the underlying microstates $\{s(\bx)\}$ visited during the 
course of a single microcanonical biased Monte Carlo simulation.

To sketch this procedure, we consider the constrained microcanonical density of states 
$g(E,o)$, which is formally written as 
\begin{eqnarray}
g(E,o)=\!\!\!\sum_{\{s(\bx)\}}\!\!\!\delta\left(o-\calO[\{s(\bx)\}]\right)\delta\left(E-\calH[\{s(\bx)\}]\right).  
\end{eqnarray}
The corresponding (conditional) probability to find the value $o$ of $\calO[\{s(\bx)\}]$ at total energy $E$ is 
\begin{eqnarray}
P(o|E)=\frac{g(E,o)}{\sum_{o'}g(E,o')}=\frac{g(E,o)}{g(E)}\propto h(o,E),  
\end{eqnarray}
where $h(o,E)$ denotes a two-dimensional histogram recorded during the course of the simulation.
But, according to the rules of conditional probabilities, 
this precisely implies that the \emph{canonical} probability to find the value $o$ of $\calO[\{s(\bx)\}]$ at inverse temperature $\beta$ is
\begin{eqnarray}
P(o|\beta)=\sum_{E}P(o|E)P(E|\beta)\propto \sum_E h(o,E)P(E|\beta).
\end{eqnarray}
To obtain this probability at any given temperature from a microcanonical Monte Carlo simulation biased by the predetermined 
density of states $g(E)$, we thus only need to reweight the recorded histograms $h(o,E)$ by the known function $P(E|\beta)$.
The desired constrained free energy density
\begin{eqnarray}
\!\!f(\beta,o)=
-\frac{1}{N\beta}\ln\!\!\sum_{\{s(\bx)\}}\!\!\delta\left(o-\calO[\{s(\bx)\}]\right)e^{-\beta\calH[\{s(\bx)\}]}  
\end{eqnarray}
can be recast in a similar way, since
\begin{eqnarray}
\lefteqn{\sum_{\{s(\bx)\}}\!\!\delta\left(o-\calO[\{s(\bx)\}]\right)e^{-\beta\calH[\{s(\bx)\}]}    
}\nonumber\\&=&
\sum_{E}e^{-\beta E}\sum_{\{s(\bx)\}}\!\!\delta\left(o-\calO[\{s(\bx)\}]\right)\delta\left(E-\calH[\{s(\bx)\}]\right)    \nonumber
\\&=&Z(\beta)\sum_{E}P(E|\beta)g(E,o)=Z(\beta)P(o|\beta),
\end{eqnarray}
and is thus (up to an unimportant constant) given by
\begin{eqnarray}
f(\beta,o)=-\frac{1}{N\beta}\ln P(o|\beta).  
\end{eqnarray}

\section{Microcanonical Results}

\begin{figure}[tb]
  \centering
\includegraphics[scale=0.65]{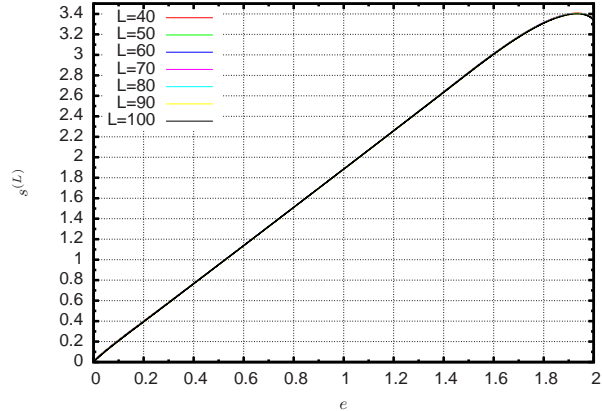}
  \caption{Microcanonical entropy densities $s^{(L)}(e)$ for the $q=30$ Potts model in $d=2$ dimensions with periodic boundary conditions
for various (but indistinguishable) system sizes.}
  \label{fig:nckccnlcnlcnlcnqlnclnclqnqlcnql}
\end{figure}

\begin{figure}[tb]
  \centering
\includegraphics[scale=0.65]{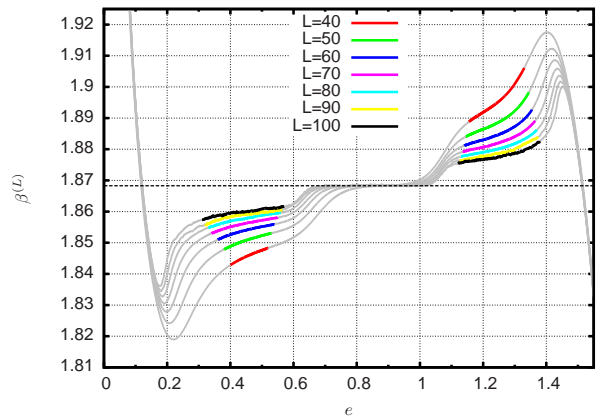}
  \caption{$\beta^{(L)}(e)$ with energy density ranges of droplets and bubbles marked;
the horizontal line displays $\beta_0 \approx 0.86829$.}
  \label{fig:klomkmfkwfn}
\end{figure}

\begin{figure}[tb]
  \centering
\includegraphics[scale=0.65]{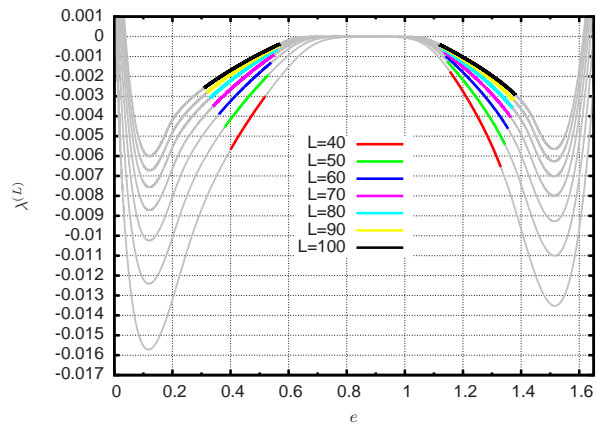}
  \caption{$\la^{(L)}(e,\beta_0)$ with energy density ranges of droplets and bubbles marked}
  \label{fig:gvonebhicwekedejdqej}
\end{figure}

We have conducted a series of Landau-Wang simulations followed by weighted Monte Carlo production runs 
of a $2d$ square lattice Potts model of size $N=L^2$ to determine the density of states $g^{(L)}(E)$ 
and thus the entropy density $s^{(L)}(e)=N^{-1}\ln g^{(L)}(E/N)$ and the microcanonical temperature $\beta^{(L)}(e)=ds^{(L)}(e)/de$.
Since we are interested in the development of phase separation in small to finite system sizes, we carried out 
simulations for linear sizes $L=40,50,\dots 100$. For the $q=10$ Potts model, the resulting signs of phase separation were not observed
to be very pronounced. However, increasing $q$ to $q=30$ clearly revealed the expected convex intruder. 
But even then, from merely looking at the entropy density $s^{(L)}(e)$ 
it is virtually impossible to detect the delicate features appearing at finite system sizes that we are interested in
(cf.\ Figure \ref{fig:nckccnlcnlcnlcnqlnclnclqnqlcnql}).
However, numerically taking the derivative of $s^{(L)}(e)$ with respect to $e$, we obtain the microcanonical inverse temperature $\beta^{(L)}(e)$
which provides a detailed view of the delicate substructures hidden in $s^{(L)}(e)$
(cf.\ Figure \ref{fig:klomkmfkwfn}). Inspection of the resulting curves gives a first hint on the quality of our simulation data,
as one should take into account that numerically differentiating potentially noisy data should greatly magnify any statistical 
irregularities and errors in such data.
To resolve the convex intruder in the original entropy data $s^{(L)}(e)$, it turns out to be convenient to consider the auxiliary dimensionless function
\cite{TaylorPaulBinder_PRE79_2009}
\begin{eqnarray}
\Lambda^{(L)}(E,\beta)/N=\la^{(L)}(e,\beta):=\beta e-s^{(L)}(e).  
\label{eqn:djxmqkxmqxmqklxmqklxmqlxnqjlxnqxnql}
\end{eqnarray}
Of course, $\Lambda^{(L)}(e,\beta)$ coincides up to a constant 
with the logarithm of the canonical energy probability function $P^{(L)}(E|\beta)$ at inverse temperature $\beta$.
However, for our present purposes we may regard $\Lambda^{(L)}$ as a finite size ``Landau potential'', i.e.~an incomplete Legendre transform of the microcanonical entropy,
and compare its features to those of canonical Landau potentials (cf.~\cite{TroesterDellagoSchranz_PRB72_2005,Troester_PRB76_2007}). 
Thus, let us tune the parameter $\beta$ to values near the bulk inverse transition temperature $\beta_0$ and analyze the resulting shape of
$\la^{(L)}(e,\beta)$ as a function of $e$. As expected, for such temperatures  $\la^{(L)}(e,\beta)$ resembles a somewhat distorted double-well shape
(cf.~Figure \ref{fig:gvonebhicwekedejdqej}) with two pronounced minima at energy densities $e_c=E_c/N$ and $e_v=E_v/N$ (the subscripts 
``c'' and ``v'' correspond to ``condensed'' and ``vapor'') separated by
a large ``Landau free energy barrier'' with a ``flat'' central plateau of practically constant and vanishing or at least
quite small slope. We identify $e_c$ and $e_v$ with the equilibrium energy densities of the bulk ``condensed'' (ordered) and ``vapor'' (disordered) 
phases, while the thermodynamics of their possible coexistence configurations in encoded in the features of the potential well between them.    
The flat central region of the potential, which signals phase separated configurations with a slab-like interface geometry \cite{TroesterDellagoSchranz_PRB72_2005}, 
 is, of course, also reflected in the central linear section of $\beta^{(L)}(e)$ at the level of $\beta^{(L)}(e)\approx\beta_0$
(cf.~Figure \ref{fig:klomkmfkwfn}). 

Apparently, Figure \ref{fig:gvonebhicwekedejdqej} also illustrates the dilemma of defining a ``proper'' finite size transition temperature
for a system with a highly degenerate low energy domain structure. On the one hand,
one could naively try to adjust $\beta$ to such a value that both minima of $\la^{(L)}(e,\beta)$ are of equal height. 
This choice precisely corresponds to the ``equal height rule'' for $P^{(L)}(E|T)$. However, at such a temperature, one observes a noticeable slope
in the central ``flat region'' of $\la^{(L)}(e,\beta)$, which signals that phase coexistence is not well established. 
In a plot of the quasi-Gaussian function
$P^{(L)}(E|T)$ this and other delicate features outside the peak regions do not give themselves away to the naked eye, since they are
exponentially suppressed.
On the other hand, choosing the inverse temperature $\beta$ to agree with the ratio-of-weights temperature with $L\to\infty$,
which, as discussed above, converges exponentially fast to the exactly known inverse bulk transition temperature
\cite{Baxter_JPC_1973}
\begin{eqnarray}
\beta_0 =\ln(1+\sqrt{q})\stackrel{(q=30)}{=}1.86829,
\end{eqnarray} 
the flat central region of $\la^{(L)}(e,\beta_0)$ is found within numerical precision to be horizontal, i.e.~with vanishing slope, 
but now one notices a pronounced difference in height between the two minima
at energy densities $e_c,e_v$, which diminishes with growing system size. Numerically, this height difference is seen to approach 
\begin{eqnarray}
\la^{(L)}(e_v,\beta_0)-\la^{(L)}(e_c,\beta_0)\approx N^{-1}\ln q.  
\label{eqn:jdnqjnwlswklssmwklslwksm}
\end{eqnarray}
Physically we can interpret these findings as follows. Suppose that precisely at the inverse transition temperature $\beta_0$
the system  initially starts out in an ordered equilibrium state. Then, the 
probability to generate a fluctuation yielding a mixed state where half of the available volume is turned into a disordered
state separated from the ordered part by a straight interfacial line, should differ from the ``inverse'' probability to produce the same state
starting from the disordered state by a factor of $q$, since in the latter case $q$ possible ordered states are available, while only one disordered
configuration can be formed in the former one. Taking the logarithm of the fraction of these probabilities then produces (\ref{eqn:jdnqjnwlswklssmwklslwksm}).
This completes the picture, since, as could have been anticipated from the practically perfect 
Gaussian nature of the two peaks in $P(E|1/\beta_0)$, the ratio-of-weights and ratio-of-heights temperatures 
are found to numerically agree for practical purposes.
Without going into the details we also note that the equal height temperature can also be shown to coincide with the inverse temperature found 
by imposing a microcanonical Maxwell construction, i.e.~chosing the value of $\beta$ for which suitable defined
areas obtained from integrating $\beta(E)$ between $E_c$ and $E_v$ coincide \cite{Janke_NPB63_1998}.

At this point, a few additional comments concerning the true nature of the above ``Landau potential'',
the non-monotonous behaviour of $\beta^{(L)}(e)$ and the resulting appearance of branches of ``negative specific heat'' are in order.
In fact, in the thermodynamic limit, $\beta (e)=\lim_{L\to\infty}\beta^{(L)}(e)$ indeed decreases with $e$ up to $e=e_c$,
stays constant at $\beta(e)=\beta_0$ up to $e=e_v$, and then decreases
further, as it should be: no trace of any metastable states (or even ``unstable
states'') is left in the $\beta(e)$ curve. Of course, this must be so: apart
from statistical errors, Monte Carlo simulations yield the equilibrium statistical
mechanics of any such model Hamiltonian exactly, and metastable or unstable curves cannot
be the output of exact calculations in the framework of equilibrium statistical mechanics.
So for $L \rightarrow \infty$ the minimum position $e_{\rm min}^{(L)}$ of $\beta^{(L)}(e)$ moves towards $e_c$ (and its depth
vanishes); similarly, the  maximum position $e_{\rm max}^{(L)}$  moves towards $e_v$ (and
its height, relative to $\beta_0$, vanishes as well). It is interesting to recall
the physical significance of these extrema: for $e_c \leq e  \leq e_{\rm min}^{(L)}$ the finite
system is still homogeneous, and the minimum is the signature of the first appearance of
a ``bubble'' of the disordered phase within the otherwise homogeneously ordered phase
(cf. Figure \ref{fig:delddmedmekmdekmekdmekmsnnsnlsn}), while the maximum is the signature of the first appearance of a
``droplet'' of the ordered phase within the otherwise homogeneous disordered phase
(Figure \ref{fig:delddmedmekmdekmekdmekm}). As long as such ``heterophase fluctuations'' are absent, finite size
effects are small in the curve $\beta^{(L)}(e)$ shown in Figure \ref{fig:klomkmfkwfn}; the strong finite
size effects in between $e_{\rm  \min}^{(L)}$ and $e_{\rm max}^{(L)}$ are due to interfacial
contributions to the ``Landau potential'' (Figure \ref{fig:gvonebhicwekedejdqej}), which are of relative order
$1/L$ in Figure \ref{fig:gvonebhicwekedejdqej}. Thus, the branch of negative ``specific heat'' resulting from Figure \ref{fig:klomkmfkwfn} in
the region where $\beta^{(L)}(e)$ is an increasing function of $e$ is not at all an
unphysical result, but simply reflects the importance of interfacial free energies in finite
microcanonical systems. In addition, our use of the nomenclature ``Landau potential'' merely
refers to the incomplete character of the Legendre transform (\ref{eqn:djxmqkxmqxmqklxmqklxmqlxnqjlxnqxnql}),
but should not mislead the reader to confuse this potential with ``Landau potentials'' of similar appearance
as they are constructed in mean-field theory. In fact, our potential $\la^{(L)}(e,\beta)$, whose information content 
is, after all, identical to that of the full microcanonical entropy density $s^{(L)}(e)$, describes the thermodynamics
of two-phase coexistence in an inhomogeneous finite system without any approximation, and thus is conceptually quite different from a 
mean-field potential, which is constructed under the implicit constraint that the system is in a homogeneous phase 
throughout, whereas such states are thermodynamically unstable in reality.

\begin{figure}[tb]
  \centering
\includegraphics[scale=0.65]{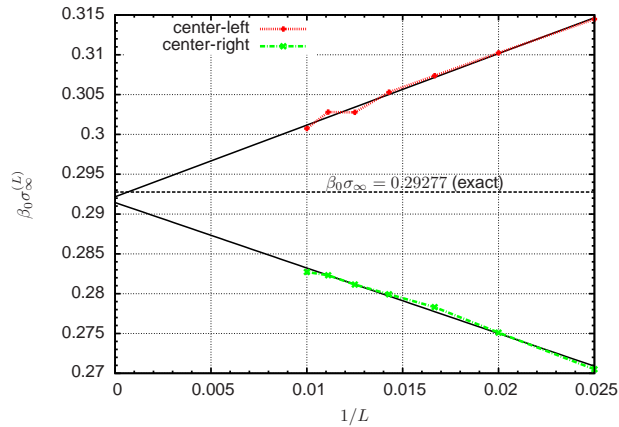}
  \caption{Finite size extrapolation (\ref{eqn:xnnjxdqkxqkxqklxqnxqnxq}) of reduced interface tension for planar interface.}
  \label{fig:klomkmfkwfnmkwmfkmkmw}
\end{figure}

We can gain confidence in the overall correctness and general quality of our data by comparing the exact value 
$\beta_0\sig=0.29277$ of the reduced $q=30$ planar interface tension as calculated from (\ref{eqn:bnchebeidqednqjdqddbnqw})
to the one obtained by a finite size extrapolation of our data.
In fact, as there are two minima of $\la^{(L)}(e,\beta_0)$ at energy densities $e_c^{(L)},e_v^{(L)}$ whose values differ by $\sim\ln 30/N$ as discussed
above, there are two corresponding sets of data $\{\la^{(L)}(e^{(L)}_{\mathrm{max}},\beta_0)-\la^{(L)}(e^{(L)}_c,\beta_0)\},\{\la^{(L)}(e^{(L)}_{\mathrm{max}},\beta_0)-\la^{(L)}(e_v^{(L)},\beta_0)\}$,
corresponding to the difference between the central barrier $\{\la^{(L)}(e^{(L)}_{\mathrm{max}},\beta_0)\}$ taken at some energy density
$e^{(L)}_{\mathrm{max}}\approx(e^{(L)}_c+e^{(L)}_v)/2$ and the left and right minima $\{\la^{(L)}(e^{(L)}_c,\beta_0)\}$, $\{\la^{(L)}(e^{(L)}_v,\beta_0)\}$, respectively
(see Figure \ref{fig:gvonebhicwekedejdqej}). A standard finite size scaling extrapolation of these data to $L\to\infty$ 
in the form
\begin{eqnarray}
\beta_0\sig_{c,v}(L)=\beta_0\sig_{c,v}(\infty)+\mathrm{const}/L,
\label{eqn:xnnjxdqkxqkxqklxqnxqnxq}  
\end{eqnarray}
which is displayed in Figure \ref{fig:klomkmfkwfnmkwmfkmkmw}, gives the two values 
$\beta_0\sig_{c}(\infty)=0.292168$ and $\beta_0\sig_{v}(\infty)=0.291441$ for the left and right difference, respectively,
whose average  $\beta_0(\sig_{c}(\infty)+\sig_{v}(\infty))/2=0.291805$ differs by less than $0.4\%$ from the exact value
$\beta_0\sig_{\infty}=0.29276$ computed from Formula (\ref{eqn:bnchebeidqednqjdqddbnqw}).  

From (\ref{eqn:ndjswnqlwqlndwqlklndqndql}), (\ref{eqn:hbebebeekkeekxc})
and (\ref{eqn:a_ndebnefwewefwebf})-(\ref{eqn:c_ndebnefwewefwebf}) it is now straightforward to compute
$\sig^{(L)}(R_s)$ and $R_s$ for any prescribed total energy density $e$. Note, however, that these formulae 
are only valid for a spherical phase separation geometry. The corresponding approximate density regions within which 
one may expect states which on average resemble spherical droplets and bubbles to dominate 
in the sampled microscopic system configurations may be found by visually inspecting
the slopes of $\la^{(L)}(e,\beta_0)$ and $\beta^{(L)}(e)$ in Figures \ref{fig:klomkmfkwfn} and \ref{fig:gvonebhicwekedejdqej}, respectively, and
cross-checking these ranges by examining corresponding snapshots taken during the course of the simulation.
The total energy density regions for which we may expect the appearance of spherical droplets and bubbles are indicated in colour in
Figures \ref{fig:klomkmfkwfn} and \ref{fig:gvonebhicwekedejdqej}. 

\begin{figure}[tb]
  \centering
\includegraphics[scale=0.65]{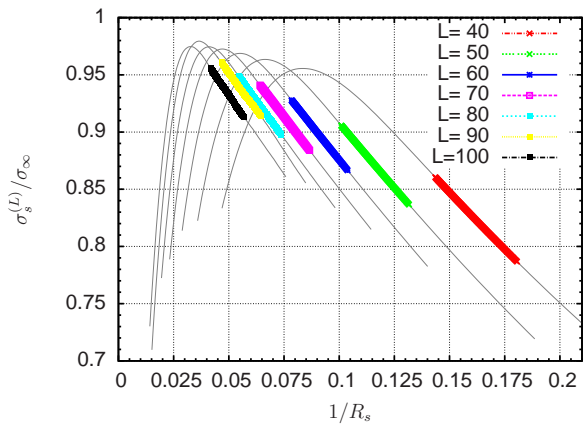}
  \caption{Normalized interface tension $\sig(R_s)/\sig_\infty$ of droplets. The ranges of radii, for which the curves actually
describe spherical droplets are marked in colour.}
  \label{fig:dropklomkmfkwfnmkwmfkmkmwjnejfnfn}
\end{figure}

\begin{figure}[tb]
  \centering
\includegraphics[scale=0.65]{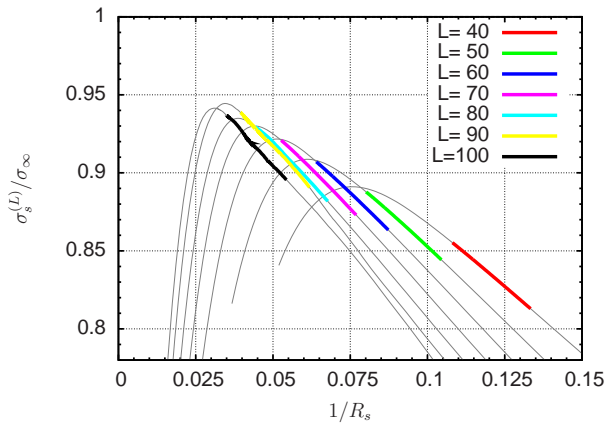}
  \caption{Normalized interface tension $\sig(R_s)/\sig_\infty$ of bubbles. The ranges of radii, for which the curves actually
describe spherical bubbles are marked in colour}
  \label{fig:bubbklomkmfkwfnmkwmfkmkmwjnejfnfn}
\end{figure}

Our results for the interface tension $\sig^{(L)}(R_s)$ at the surface of tension are gathered in 
Figures \ref{fig:dropklomkmfkwfnmkwmfkmkmwjnejfnfn} and \ref{fig:bubbklomkmfkwfnmkwmfkmkmwjnejfnfn}. In correctly interpreting these results,
it is quite important to understand that they have been obtained from (\ref{eqn:ndjswnqlwqlndwqlklndqndql}), (\ref{eqn:hbebebeekkeekxc})
and (\ref{eqn:a_ndebnefwewefwebf})-(\ref{eqn:c_ndebnefwewefwebf}) under the assumption of spherical geometry. The corresponding ranges of inverse radii
for which one can expect this assumption to be valid have been marked in colour in these figures. Outside of these ranges, the data do not 
accurately describe a physical interface tension, but merely serve as a guide to the eye. 

Looking at these results, one instantly notices the large finite size effects, manifesting themselves in the considerable offsets between
the consecutive considered $L$-values, which strikingly fail to collapse onto a common ``master curve''. 
Currently, we find it difficult to understand the origin of this behaviour. For the planar interface tension,
which is described by our data quite accurately as discussed above,
strong but regular finite size effects are indeed expected. They may be heuristically understood in terms of the $L$-dependent truncation of the wave vector spectrum 
of capillary waves running parallel to the interface. 
However, for a spherical bubble or droplet, identical radii $R$ should yield
identical values of the interface tension, once the surrounding box has been chosen large enough to kill finite correlation length effects,
which are however expected to be vanishingly small for a strong first order phase transition. 

At the moment we do not have a clear explanation of these strong finite size effects which prevent us from a further meaningful analysis
of the curvature dependence of $\sig(R)$. In similar approaches to study the interface tension of curved interfaces in a truncated 3d Lennard-Jones
fluid \cite{TrosterOettelBlockVirnauBinder_inpreparation} and a 3d fcc lattice gas model \cite{TrosterBinder_PRL_2011_accepted}, 
we also find certain finite size effects, but they are much less pronounced than those of the present case.
However, we have also observed finite size effects of comparable size in computing the interface tension from canonical simulations of a 2d Ising model.
Thus, we believe that the large magnitude of the finite size effects has nothing to do with our microcanonical approach, but is rather related to the 
fact that both systems are two-dimensional. With growing linear system size $L$, the gaps between consecutive $\sig^{(L)}(R_s)$-curves recorded in Figures 
\ref{fig:bubbklomkmfkwfnmkwmfkmkmwjnejfnfn} and \ref{fig:bubbklomkmfkwfnmkwmfkmkmwjnejfnfn} obviously diminish, so these curves are expected to eventually
collapse onto a single ``master curve'' for large $L$ and $R_s$. On the other hand, for $L\ge110$ we report that our Monte Carlo production runs failed to be
sufficiently ergodic, indicating large residual entropic barriers with respect to ``hidden'' observables beyond our one-dimensional energy-based sampling,
while at the same time the barriers observed in $\Gamma^{(L)}(E)$ had already risen to a value of $~60$.   
To study much larger systems would thus require to overcome these additional hidden barriers, presumably by employing much more elaborate sampling techniques than the ones we are using here (see e.g.~\cite{MartinMayor_PRL98_2007}
or \cite{Bauer_JStMech_2010} for promising approaches). However, in our current work we are not interested in exceedingly large droplet sizes and 
their accompanying huge entropy and free energy barriers. Rather, out intention is to focus on the behavior of droplet and bubbles of moderate size, as this
is the only regime that is of practical relevance for nucleation related questions. In any case, the origin and nature of the encountered finite size effects must currently be left to further study.
 
\section{Conclusions}

In this paper, we have addressed the investigation of phase coexistence of systems with a
more-component parameter in the context of computer simulations, which necessarily involve
systems of finite size. Such simulations of phase coexistence often are done with the motivation
to extract information on the interfacial tension of flat and curved interfaces. While for
systems with a scalar (i.e.~one-component) order parameter this problem is normally considered
in the grand-canonical and canonical ensemble of statistical mechanics, we have given a concise
discussion of this approach to show that its extension to the multi-component case is formally
possible but practically unfeasible. We then have presented, as an alternative, a microcanonical
approach based on the number of states $g(E,V)$ of energy $E$ for a system having a finite
volume $V$. In the entropy versus energy curve $S(E)$ for the finite system there is a convex
intruder (Figure \ref{fig:iiibbfehbhebhfwevgwevwvw}), and the idea we follow in the present paper is to carefully analyze
this intruder as a function of system size, in order to extract information on interfacial tensions.
We exemplify our approach for the two-dimensional $q$-state Potts model with a large number of
states ($q=30$), proposing also an extension of the heat bath algorithm from the canonical to the
microcanonical ensemble. From these simulations we obtain very precise information on $S(E)$ and also
an effective potential $\la^{(L)}(e, \beta_0$) per lattice site, $e$ being the energy density and
$\beta_0$ the inverse temperature where in the thermodynamic limit ($V = L^2 \rightarrow \infty$)
the first-order transition from the ordered phase to the disordered phase occurs (Figure \ref{fig:gvonebhicwekedejdqej}).
Also the derivative $\beta^{(L)}(e)= d \la^{(L)}(e)/de$ is obtained with meaningful accuracy (Figure \ref{fig:klomkmfkwfn}). We
have shown that the loop in such $\beta^{(L)}(e)$ vs.~$e$ curves has nothing to do with the
``van der Waals-like'' loop of mean-field theories: in the latter, such loops describe a path of
homogeneous states connecting the two phases between which the transition occurs; in reality, our
loops (Figure \ref{fig:klomkmfkwfn}) reflect two phase coexistence in finite systems, all parts of the loop describe
full stable thermal equilibrium; any interpretation in terms of metastable or unstable states would be
completely misleading. There is nothing mysterious about the ``negative specific heat'' that often is
attributed to such loops - the whole loop just reflects interfacial effects, just as the ''hump'' in between the two
minima of the ``Landau potential'' in Figure \ref{fig:gvonebhicwekedejdqej}; all these features disappear proportional to
$1/L$ in the limit $ L \rightarrow \infty$, and the correct horizontal parts in between $e_c$ and
$e_v$ remain, as it should be. Thus, one should not be mislead by mean-field concepts when discussing
first-order phase transitions in finite system in the microcanonical ensemble.

We have found that in the flat region in the center of Figure \ref{fig:gvonebhicwekedejdqej} the data allow an accurate
estimation of the interfacial tension of flat interfaces between ordered and disordered phase
(Figure \ref{fig:klomkmfkwfnmkwmfkmkmw}), although also in this case finite size effects are clearly rather pronounced, and
an extrapolation to $L \rightarrow \infty$ is mandatory. However, the analysis of the ascending parts
of $\lambda^{(L)}(e, \beta_0)$ in Figure \ref{fig:gvonebhicwekedejdqej} in terms of the radius-dependent interface tension of droplets
(Figure \ref{fig:dropklomkmfkwfnmkwmfkmkmwjnejfnfn}) and bubbles (Figure \ref{fig:bubbklomkmfkwfnmkwmfkmkmwjnejfnfn}) is more subtle: 
again huge finite size effects occur, and it is
not possible at fixed radius $R_s$ to extrapolate to $L \rightarrow \infty$, because due to the droplet
(bubble) evaporation/condensation transitions droplets at fixed radius $R_s$ are only stable in a rather
restricted range of $L$. While naively one could expect that different choices of $L$ yield mutually compatible
results for $\sigma(R_s)$, as approximately happens for one-component systems in $d=3$ dimensions, this is not
the case here. Of course, our analysis does not explicitly consider the fact that at a given value of $e$
and the corresponding average value of $\beta$ (Figure \ref{fig:klomkmfkwfn}) in the two-phase coexistence region at a
given value of $L$ the droplet (or bubble) is strongly fluctuating both with respect to its size and its shape
(Figures \ref{fig:delddmedmekmdekmekdmekm} and \ref{fig:delddmedmekmdekmekdmekmsnnsnlsn}). 
We assume the shape of the droplet or bubble to be spherical, otherwise the
information recorded does not suffice to extract $\sigma(R_s)$. Future work along such lines must analyze this
problem of droplet (bubble) fluctuations more closely, possibly by recording additional observables related to the droplet or 
bubble, to allow estimation of $\sigma(R_s)$ within reasonable error limits. In fact, in $d=3$ the
fluctuations are found to be indeed much less pronounced, and -- at last in the one-component case --
meaningful results for $\sigma(R)_s$ are accessible \cite{TrosterOettelBlockVirnauBinder_inpreparation,TrosterBinder_PRL_2011_accepted}. 

\ack

A. Tr\"oster acknowledges support by the Austrian Science Fund (FWF): P22087-N16 and is grateful to Prof.~J.~Henderson for 
enlightening emails and to D.~Reith for valuable computer assistance.
K. Binder received support from the Deutsche Forschungsgemeinschaft (DFG) under grant No SPP 1296/BI 314/19-2. 

\section{References}

\bibliographystyle{/home/troester/Desktop/LaTeX/bib/prsty}
\bibliography{PottsHighQ2011IOP.bib}
\end{document}